\documentclass[aps,prc,showpacs,twocolumn,floatfix]{revtex4}

\usepackage{amsmath}
\usepackage{bm}
\usepackage{mathrsfs}
\usepackage{graphicx}
\usepackage{subfigure}

\renewcommand{\bar}[1]{\overline{#1}}
\providecommand{\Journal}[4] {#1 {\bf#2}, #3 (#4)}
\providecommand{\PLB}{Phys. Lett. B} %
\providecommand{\PRL}{Phys. Rev. Lett.} %
\providecommand{\PRD}{Phys. Rev. D}
\providecommand{\PRC}{Phys. Rev. C} %
\providecommand{\NPA}{Nucl. Phys. A} %
\providecommand{\EPJA}{Eur. Phys. J. A} %
\providecommand{\PR}{Phys. Rep.} %
\providecommand{\IJMPA}{Int. J. Mod. Phys. A } %
\providecommand{\HEPNP}{High Energy Phys. Nucl. Phys. } %

\begin{document}

\title{Possible $\Sigma({1\over2}^-)$ under the $\Sigma^*(1385)$ peak in  $K\Sigma^*$ photoproduction}
\author{Puze Gao, Jia-Jun Wu and B.S.~Zou}
\affiliation{Institute of High Energy Physics, CAS, P.O. Box 918(4),
Beijing 100049, China} \affiliation{Theoretical Physics Center for
Science Facilities, CAS, Beijing 100049, China}

\begin{abstract}
The LEPS collaboration has recently reported a measurement of the
reaction $\gamma n\to K^+\Sigma^{*-}(1385)$ with linearly polarized
photon beam at resonance region. The observed beam asymmetry is
sizably negative at $E_\gamma=1.8-2.4 \mathrm{GeV}$, in contrast to
the presented theoretical prediction. In this paper, we calculate
this process in the framework of the effective Lagrangian approach.
By including a newly proposed $\Sigma(J^P={1\over2}^-)$ state with
mass around 1380~MeV, the experimental data for both $\gamma n$ and
$\gamma p$ experiments can be well reproduced. It is found that the
$\Sigma({1\over2}^-)$ and/or the contact term may play important
role and deserve further investigation.
\end{abstract}

\maketitle

\section{INTRODUCTION}

With the development of accelerator facilities, strangeness
production from photon-nucleon scattering has been extensively
studied at resonance
region~\cite{clas06,clasg06,clas07,ct08,sp98,leps03,leps06,leps08,leps09}.
These experiments can provide not only more information on the
properties and interactions of the known resonances, but also clues
for the existence of some new resonances. For $K\Sigma^*(1385)$
photoproduction, high statistic data have been available only
recently. The CLAS collaboration has studied the reaction $\gamma
+p\rightarrow K^++\Sigma^{*0}(1385)$ with unpolarized photon beam at
energy $E_\gamma=1.5-4$~GeV~\cite{clasg06}. The LEPS collaboration
has reported the reaction $\gamma +n\rightarrow
K^++\Sigma^{*-}(1385)$, with a linearly polarized photon beam at
$E_\gamma=1.5-2.4$~GeV~\cite{leps09}. The high statistic data and
the polarized observables provide more information and challenges
for theoretical studies.

Theoretical investigations on $K\Sigma^*(1385)$ photoproduction
include the work by Lutz and Soyeur~\cite{lutz05}, which mainly
studies the t-channel processes, the work by D\"{o}ring, Oset and
Strottman~\cite{oset06}, where the role of $\Delta(1700)$ is
addressed, and the work by Oh, Ko, and Nakayama~\cite{oh08}, where
the roles of $N$ and $\Delta$ resonances in s-channel are stressed
and compared with the CLAS data~\cite{clasg06}.

For the approach by Oh, Ko, and Nakayama~\cite{oh08}, the total
cross section for $\gamma +p\rightarrow K^++\Sigma^{*0}(1385)$ is
calculated in the framework of gauge-invariant effective
Lagrangians. The results are in reasonable agreement with the CLAS
data~\cite{clasg06}, showing that the s-channel $N$ and $\Delta$
resonances above $K\Sigma^*$ threshold may give important
contributions to cross sections. Although this theory can well
describe the total cross section of $K\Sigma^*$ photoproduction from
CLAS as well as from LEPS experiment~\cite{leps09}, the theoretical
prediction deviates greatly from the data for linear beam asymmetry
measured by LEPS with polarized photon beam. This is an urgent
problem for theoretical studies.

From studies of baryon spectroscopy and structures, five quark
$qqqq\bar q$ components are proposed to play important roles in some
baryons~\cite{riska02,zou08}. A few years ago, Jaffe and Wilczek
have promoted a diquark-diquark-antiquark picture for the pentaquark
baryons~\cite{jaffe03}. Zhang \textit{et al.} then studied the
$J^P={1\over 2}^-$ pentaquark baryons based on this picture and
predicted a $\Sigma({1\over2}^-)$ state with mass around
1360~MeV~\cite{zhu05}. A more general pentaquark
model~\cite{riska02} without introducing explicitly diquark clusters
predicts that $\Sigma({1\over 2}^-)$ has a mass similar to
$\Lambda({1\over 2}^-)$, which is around 1405 MeV. From these two
models, one would expect a $\Sigma({1\over2}^-)$ state with mass
around 1380~MeV. Recent studies on $K^-p\to \Lambda\pi^+\pi^-$
process have shown some evidence for the existence of the
$\Sigma({1\over2}^-)$ near 1380~MeV~\cite{wu09}. In this work, we
study the $K\Sigma^*(1385)$ photoproduction processes with the
consideration of the case that a portion of $K\Sigma({1\over2}^-)$
photoproduction is mixed in.

This paper is organized as follows. In section II, the theoretical
framework is presented for the $K\Sigma^*$ and
$K\Sigma({1\over2}^-)$ photoproduction from the nucleons. In section
III, the numerical results for cross sections and the beam asymmetry
are presented and compared with the experimental data, with some
discussions. In section IV, we give the summery of this work.

\section{THEORETICAL FRAMEWORK}

The effective Lagrangian method is an important theoretical approach
in describing the various processes at resonance region. We use the
effective Lagrangians of Ref.~\cite{oh08} for $K\Sigma^*$
photoproduction, where the contact term is derived from
Ref.~\cite{hab06} to keep the amplitude gauge invariant. In the
following equations we use $\Sigma^*$ and $\Sigma$ denoting the
$\Sigma^*({3\over2}^+)$ at 1385~MeV and the $\Sigma({1\over2}^-)$
near 1380~MeV, respectively.

\subsection{$K\Sigma^*({3\over 2}^+)$ photoproduction}

For the reaction $\gamma N\to K \Sigma^*({3\over2}^+)$, the Feynman
diagrams are shown in Fig.~1, where the incoming momenta are $k$ and
$p$ for photon and nucleon, respectively, and the outgoing momenta
are $q$ and $p'$ for $K$ meson and the $\Sigma^*$, respectively.
From the investigation of Ref.~\cite{oh08}, the main contributions
come from the t-channel $K$ meson exchange, the s-channel $N$ and
$\Delta$ as well as their resonances exchange, the u-channel
$\Lambda$ (for neutral propagator only) and $\Sigma^*({3\over2}^+)$
exchange and the contact term.
\begin{figure}\label{fig:fds}
{\includegraphics[width=0.80\columnwidth]{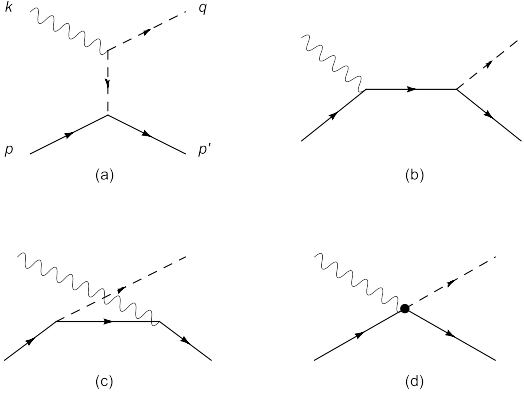}}
 \caption{Feynman diagrams for $\gamma +N\rightarrow
K+\Sigma^*({3\over2}^+)$. (a) t-channel; (b) s-channel; (c)
u-channel; (d) contact term.}
\end{figure}

For the t-channel $K$ meson exchange, the relevant effective
Lagrangians are
\begin{equation}\label{gkk}
{\cal L}_{\gamma KK}=i e A_\mu(K^-\partial^\mu K^+-\partial^\mu K^-
K^+)\,,
\end{equation}
\begin{equation}\label{knss}
{\cal L}_{KN\Sigma^*}={f_{KN\Sigma^*}\over
m_K}\partial_\mu\bar{K}{\bf\bar{\Sigma}^{*\mu}}\cdot
{\bf\tau}N+H.c.\,,
\end{equation}
with the isospin structure of $K\Sigma^*N$ coupling
\begin{equation}\label{isostr}
\bar K=(K^-,{\bar K}^0), {\bf\bar{\Sigma}}\cdot
{\bf\tau}=\left ( \begin{array}{cc} \bar{\Sigma}^0& \sqrt{2}{\bar\Sigma}^+ \\
\sqrt{2}{\bar\Sigma}^- & -{\bar\Sigma}^0
\end{array}\right ),N=\left(\begin{array}{c} p
\\n\end{array}\right),
\end{equation}
where $f_{KN\Sigma^*}$ is the coupling constant and is taken as
$f_{KN\Sigma^*}=-3.22$ from the SU(3) flavor symmetry relation
 as in Ref.~\cite{oh08}; $m_K$ is the mass for $K$ meson.

For the s-channel nucleon exchange, the effective Lagrangians for
the interactions contain Eq.~(\ref{knss}) as well as
\begin{equation}\label{gnn}
{\cal L}_{\gamma NN}=-e \overline{N}(\gamma^\mu A_\mu
Q_N-{\kappa_N\over 2M_N}\sigma^{\mu\nu}\partial_\nu A_\mu)N\,,
\end{equation}
where $Q_N$ is the electric charge (in unite of $e$), and $\kappa_N$
 denotes the magnetic moment of the nucleon.

For the s-channel spin-${3\over 2}$ and spin-${5\over 2}$ resonances
exchange, the effective Lagrangians for the interactions are
\begin{eqnarray}
{\cal L}_{\gamma NR}({3\over 2}^\pm)&=&-{ief_1\over 2M_N}{\bar
N}\Gamma_\nu^{(\pm)}F^{\mu\nu}R_\mu \nonumber\\
&&-{ef_2\over (2M_N)^2}\partial_\nu{\bar
N}\Gamma^{(\pm)}F^{\mu\nu}R_\mu + \mathrm{H.c.},\\
{\cal L}_{\gamma NR}({5\over 2}^\pm)&=&{ef_1\over (2M_N)^2}{\bar
N}\Gamma_\nu^{(\mp)}\partial^\alpha F^{\mu\nu}R_{\mu\alpha} \nonumber\\
&&-{ief_2\over (2M_N)^3}\partial_\nu{\bar
N}\Gamma^{(\mp)}\partial^\alpha F^{\mu\nu}R_{\mu\alpha} +
\mathrm{H.c.},
\end{eqnarray}
and
\begin{eqnarray}
{\cal L}_{RK\Sigma^*}({3\over 2}^\pm)&=&-{h_1\over
m_K}\partial^\alpha K{\bar\Sigma}^{*\mu}\Gamma_\alpha^{(\pm)}R_\mu\nonumber\\
&&+{ih_2\over (m_K)^2}\partial^\mu\partial^\alpha
K{\bar\Sigma}^*_\alpha\Gamma^{(\pm)}R_\mu+ \mathrm{H.c.},\\
{\cal L}_{RK\Sigma^*}({5\over 2}^\pm)&=&{ih_1\over
m_K^2}\partial^\mu\partial^\beta K{\bar\Sigma}^{*\alpha}\Gamma_\mu^{(\mp)}R_{\alpha\beta}\nonumber\\
&&-{h_2\over (m_K)^3}\partial^\mu\partial^\alpha\partial^\beta
K{\bar\Sigma}^*_\mu\Gamma^{(\mp)}R_{\alpha\beta}+ \mathrm{H.c.},
\end{eqnarray}
where $F^{\mu\nu}=\partial^\mu A^\nu-\partial^\nu A^\mu$; $R_\mu$
and $R_{\mu\alpha}$ denote the spin-${3\over 2}$ and spin-${5\over
2}$ fields, respectively, and
\begin{equation}
\Gamma_\mu^{(\pm)}=\left(\begin{array}{c} \gamma_\mu\gamma_5
\\\gamma_\mu\end{array}\right), \Gamma^{(\pm)}=\left(\begin{array}{c}
\gamma_5
\\1\end{array}\right).
\end{equation}
For $\Delta$ and $\Delta$ resonances of isospin-3/2, the effective
Lagrangians have the isospin structure
\begin{eqnarray}
\bar{K} \bar{\bf{\Sigma}}^*\cdot {\bf T}({1\over 2},{3\over 2})
\Delta=\sqrt{3}K^-\bar{\Sigma}^{*+}\Delta^{++}
-\sqrt{2}K^-\bar{\Sigma}^{*0}\Delta^{+} \nonumber\\
-K^-\bar{\Sigma}^{*-}\Delta^{0}
+\bar{K}^0\bar{\Sigma}^{*+}\Delta^{+}
-\sqrt{2}\bar{K}^0\bar{\Sigma}^{*0}\Delta^{0}\nonumber\\
-\sqrt{3}\bar{K}^0\bar{\Sigma}^{*-}\Delta^{-}.
\end{eqnarray}
We consider 3 PDG resonances in s-channel, namely, $N_{{3\over
2}^-}(2080)$, $\Delta_{{3\over 2}^-}(1940)$ and $\Delta_{{5\over
2}^+}(2000)$, which are the most prominent ones as stated in
Ref.~\cite{oh08}. The coupling constants $f_1$ and $f_2$ can be
calculated from the helicity amplitudes in PDG~\cite{PDG08} or model
predictions with Eq.(B3) of Ref.~\cite{oh08}. For $\gamma N\Delta$
coupling, we have $f_1=4.04$ and $f_2=3.87$. From the predicted
helicity amplitudes in Ref.~\cite{capstick92}, one gets $f_1=-1.25$
and $f_2=1.21$ for $\gamma pN^*(2080)$ coupling; $f_1=0.381$ and
$f_2=-0.256$ for $\gamma nN^*(2080)$ coupling; $f_1=0.39$ and
$f_2=-0.57$ for $\gamma N\Delta(1940)$ coupling, and $f_1=-0.68$,
$f_2=-0.062$ for $\gamma N \Delta(2000)$ coupling. For $\Delta
K\Sigma^*$ coupling, $h_1=2.0$ and $h_2=0$ are used from
$h_1=-f_{K\Delta\Sigma^*}/{\sqrt{3}}$ with
$f_{K\Delta\Sigma^*}=-3.46$~\cite{ohprt06}. For the resonances
coupling to $K\Sigma^*$, the coupling constants $h_1$ and $h_2$ can
be calculated from Eq.(B11)-(B18) of Ref.~\cite{oh08} from the model
predicted amplitudes $G(l)$~\cite{capstick98}. In our concrete
calculation, $h_1=0.24$ and $h_2=-0.54$ for $N^*(2080)K\Sigma^*$
coupling, $h_1=-0.68$ and $h_2=1.0$ for $\Delta(1940)K\Sigma^*$
coupling, and $h_1=-1.1$ and $h_2=0.21$ for $\Delta(2000)K\Sigma^*$
coupling are taken.

The reaction $\gamma p\rightarrow K^+\Sigma^{*0}$ contains the
u-channel $\Lambda(1116)$ exchange. The effective Lagrangians are
\begin{eqnarray}
{\cal L}_{\gamma\Lambda\Sigma^*}&=&-{ief_1\over 2M_\Lambda}{\bar
\Lambda}\gamma_\nu\gamma_5F^{\mu\nu}\Sigma^*_\mu \nonumber\\
&&-{ef_2\over (2M_\Lambda)^2}\partial_\nu{\bar
\Lambda}\gamma_5F^{\mu\nu}\Sigma^*_\mu + \mathrm{H.c.},
\end{eqnarray}
\begin{equation}
{\cal L}_{KN\Lambda}={g_{KN\Lambda}\over
M_N+M_\Lambda}\bar{N}\gamma^\mu\gamma_5\Lambda\partial_\mu
K+\mathrm{H.c.},
\end{equation}
where $f_1=4.52$, $f_2=5.63$ are obtained from decay width
$\Gamma(\Sigma^*\rightarrow\Lambda\gamma)$ and model predicted
helicity amplitudes; and $g_{KN\Lambda}=-13.24$ is estimated from
flavor SU(3) symmetry relation as in Ref.~\cite{oh08}.

For u-channel $\Sigma^*$ exchange, the effective Lagrangian for
$\gamma\Sigma^*\Sigma^*$ is
\begin{equation}
{\cal L}_{\gamma\Sigma^*\Sigma^*}=e\bar{\Sigma}^*_\mu A_\alpha
\Gamma^{\alpha,\mu\nu}_{\gamma\Sigma^*}\Sigma^*_\nu,
\end{equation}
with
\begin{eqnarray}
A_\alpha
\Gamma^{\alpha,\mu\nu}_{\gamma\Sigma^*}&=&Q_{\Sigma^*}A_\alpha\big
(g^{\mu\nu}\gamma^\alpha-{1\over
2}(\gamma^\mu\gamma^\nu\gamma^\alpha+\gamma^\alpha\gamma^\mu\gamma^\nu)\big )\nonumber\\
&&-{\kappa_{\Sigma^*}\over 2M_N}\sigma^{\alpha\beta}\partial_\beta
A_\alpha g^{\mu\nu},
\end{eqnarray}
where $Q_{\Sigma^*}$ denotes the electric charge of $\Sigma^*$ in
unit of $e$ and $\kappa_{\Sigma^*}$ denotes the anomalous magnetic
moment of $\Sigma^*$; $\kappa_{\Sigma^{*0}}=0.36$ and
$\kappa_{\Sigma^{*-}}=-2.43$ are taken from the quark model
prediction~\cite{Lich77}.

For each vertex of these channels, a form factor is attached to
describe the off-shell properties of the amplitudes. For t-channel
$K$ meson exchange, we adopt the form factor~\cite{oh08}
\begin{equation}\label{FM}
F_M={\Lambda_M^2-m_K^2\over \Lambda_M^2-q_t^2}\, ,
\end{equation}
where $q_t=k-q$ is the 4-momentum transfer in t-channel. For the
s-channel and u-channel processes, we adopt the form factor
\begin{equation}\label{FB}
F_B(q_{ex}^2,M_{ex})=\left({n\Lambda_B^4\over
n\Lambda_B^4+(q_{ex}^2-M_{ex}^2)^2}\right )^n\, ,
\end{equation}
where the $q_{ex}$ and $M_{ex}$ are the 4-momentum and the mass of
the exchanged hadron, respectively. For s-channel $N$ and $\Delta$
exchange and the u-channel processes, we take $n=1$ as in
Ref.~\cite{oh08}; for s-channel resonances exchange,
$n\rightarrow\infty$ is taken to obtain the Gaussian form for the
form factors \cite{oh08}
\begin{equation}\label{FBR}
F_B(q_s^2,M_R)|_{n\rightarrow\infty}=\exp\Big(-{(q_s^2-M_R^2)^2\over
\Lambda_B^4}\Big)\,.
\end{equation}
The cutoff parameters $\Lambda_M$ and $\Lambda_B$ were taken to be
0.83 and 1.0 GeV, respectively, in Refs.~\cite{leps09,oh08}, by
fitting to the $\gamma p\to K^+\Sigma^{*0}$ data.

The contact term illustrated with Fig.~1(d) serves to keep the full
amplitude gauge invariant. The contact currents for $K\Sigma^*$
photoproduction are related to the Lagrangian of Eq.(\ref{knss}).
For the process $\gamma p\rightarrow K^+\Sigma^{*0}$, we adopt the
contact current~\cite{hab06,oh08}
\begin{equation}
M_c^{\mu\nu}=ie{f_{KN\Sigma^*}\over m_K}(g^{\mu\nu}f_t-q^\mu C^\nu),
\end{equation}
where $C^\nu$ is expressed as
\begin{eqnarray}\label{cnup}
C^\nu=-(2q-k)^\nu{f_t-1\over
t-m_K^2}\big(1-h(1-f_s)\big )\nonumber\\
-(2p+k)^\nu{f_s-1\over s-M_N^2}\big(1-h(1-f_t)\big )\,.
\end{eqnarray}
Here the Lorenz index $\mu$ and $\nu$ couple to that of $\Sigma^*$
and the photon respectively; $f_t=F_M^2$ and $f_s=F_B^2(s,M_N)$ are
form factors squared, and $t=q_t^2$ and $s=q_s^2$ are squared
momentum transfer for t- and s-channel; $h$ is a parameter to be
fitted to experiments, and $h=1$ is used in Ref.\cite{oh08}. From
this contact term, we can check that the total amplitude is gauge
invariant. For the process $\gamma n\rightarrow K^+\Sigma^{*-}$, the
contact current is~\cite{hab06}
\begin{equation}
M_c^{\mu\nu}=ie\sqrt{2}{f_{KN\Sigma^*}\over m_K}(g^{\mu\nu}f_t-q^\mu
C^\nu),
\end{equation}
with
\begin{eqnarray}\label{cnun}
C^\nu=-(2q-k)^\nu{f_t-1\over t-m_K^2}\big(1-h(1-f_u)\big )\nonumber\\
+(2p'-k)^\nu{f_u-1\over u-M_{\Sigma^*}^2}\big(1-h(1-f_t)\big ).
\end{eqnarray}
Where $f_u=F_B^2(u,M_\Sigma^*)$ is form factor squared, and
$u=q_u^2$ is squared momentum transfer for u-channel.

For the propagators, we use $1/(q_t^2-m_K^2)$ for t-channel $K$
meson exchange. For the propagator of a baryon with mass $m$ and
4-momentum $p$, we use $ {\not \! p}+m\over p^2-m^2 $ for spin-1/2
propagator;
\begin{equation}
{{\not\! p}+m\over p^2-m^2}\Big(-g^{\mu\nu}+{
\gamma^\mu\gamma^\nu\over 3}+{\gamma^\mu p^\nu -\gamma^\nu
p^\mu\over 3m}+{2p^\mu p^\nu\over 3m^2}\Big)
\end{equation}
for spin-3/2 propagator; and
\begin{equation}
{{\not\! p}+m\over p^2-m^2}S_{\alpha\beta\mu\nu}(p,m)
\end{equation}
 for spin-5/2 propagator, where
\begin{eqnarray}
S_{\alpha\beta\mu\nu}(p,m)={1\over 2}({\bar g}_{\alpha\mu}{\bar
g}_{\beta\nu}+{\bar g}_{\alpha\nu}{\bar g}_{\beta\mu})-{1\over
5}{\bar g}_{\alpha\beta}{\bar g}_{\mu\nu}\nonumber\\
-{1\over 10}({\bar\gamma}_\alpha{\bar\gamma}_\mu{\bar
g}_{\beta\nu}+{\bar\gamma}_\alpha{\bar\gamma}_\nu{\bar
g}_{\beta\mu}+{\bar\gamma}_\beta{\bar\gamma}_\mu{\bar
g}_{\alpha\nu}+{\bar\gamma}_\beta{\bar\gamma}_\nu{\bar
g}_{\alpha\mu}),
\end{eqnarray}
with
\begin{eqnarray}
{\bar g}_{\mu\nu}=g_{\mu\nu}-{p_\mu p_\nu\over m^2},\nonumber\\
{\bar\gamma}_\mu=\gamma_\mu-{p_\mu\over m^2}{\not\! p}.
\end{eqnarray}
For the s-channel resonances with sizeable width $\Gamma$, we
replace the denominator $1\over p^2-m^2$ in the propagators by
$1\over p^2-m^2+im\Gamma$, and replace $m$ in the rest of the
propagators by $\sqrt{p^2}$.

\subsection{$K\Sigma({1\over 2}^-)$ photoproduction}

Since five quark components for baryons may exist, there may be a
$\Sigma({1\over 2}^-)$ that has a large probability of five quark
structure with mass near 1380 MeV as models
predict~\cite{riska02,zhu05}. Both $\Sigma^*({3\over 2}^+)$ and
$\Sigma({1\over 2}^-)$ decay strongly to $\Lambda\pi$, which are
detected by experiments. Thus we consider that $K\Sigma({1\over
2}^-)$ photoproduction may also contribute to the measured
 cross sections. We constrain our study to processes
with $s$ and $p$ wave hadronic vertices, since the contributions
from higher waves are relatively suppressed. We also neglect the
contributions from some resonances either for lack of information on
the couplings or the couplings are small. Thus the main
contributions to $K\Sigma({1\over2}^-)$ photoproduction are from the
t-channel $K$ meson exchange, the s-channel $N$ exchange, the
u-channel $\Sigma({1\over2}^-)$ exchange (and $\Lambda$ exchange for
$\gamma p\rightarrow K^+ \Sigma^0({1\over2}^-)$) and the contact
term. The Feynman diagrams are the same as Fig. 1. The effective
Lagrangian for $KN\Sigma({1\over 2}^-)$ coupling can be expressed
as~\cite{ohprt06}
\begin{equation}\label{kns}
{\cal L}_{KN\Sigma}=-i g_{KN\Sigma}\bar{K}{\bf\bar{\Sigma}}\cdot
{\bf\tau}N+\mathrm{H.c.}\,,
\end{equation}
where the coupling constant $g_{KN\Sigma}$ is to be fitted to
experiments; The isospin structure is the same as Eq.(\ref{isostr}).
The effective Lagrangians for $\gamma KK$ and $\gamma NN$ are
described in Eq.(\ref{gkk}) and Eq.(\ref{gnn}), respectively. For
$\gamma\Sigma({1\over 2}^-)\Sigma({1\over 2}^-)$ vertex, the
effective Lagrangian can be expressed as~\cite{ohprt06}
\begin{equation}
{\cal L}_{\gamma \Sigma\Sigma}=-e \overline{\Sigma}(\gamma^\mu A_\mu
Q_\Sigma -{\kappa_\Sigma\over 2M_N}\sigma^{\mu\nu}\partial_\nu
A_\mu)\Sigma\,,
\end{equation}
where $Q_\Sigma$ is the electric charge (in unite of $e$), and
 $\kappa_\Sigma$ denotes the anomalous magnetic moment for
$\Sigma({1\over 2}^-)$, and we take the predicted values
$\kappa_{\Sigma^0}=-0.43$ and $\kappa_{\Sigma^-}=-1.74$ from the
diquark model~\cite{zhu05}. For the $\gamma\Lambda\Sigma^0({1\over
2}^-)$ coupling in the $\gamma p\rightarrow K^+\Sigma^0$ process,
the effective Lagrangian is expressed as~\cite{ohprt06}
\begin{equation}
{\cal L}_{\gamma \Lambda\Sigma}={eg_{\gamma\Lambda\Sigma}\over
4(M_\Lambda+M_{\Sigma})}{\bar\Sigma}\gamma_5\sigma_{\mu\nu}\Lambda
F^{\nu\mu}+\mathrm{H.c.}\,,
\end{equation}
where we take $g_{\gamma\Lambda\Sigma}=1.16$. Note that although
this coupling has some uncertainty, it is
largely suppressed and has very small effect in this process.

For each vertex in the Feynman diagrams, a form factor is attached.
Similar to the $K\Sigma^*({3\over 2}^+)$ photoproduction processes,
we adopt the form factor as in Eq.(\ref{FM}) for t-channel $K$ meson
exchange, and Eq.(\ref{FB}) for s-channel and u-channel processes.

The contact term for $K\Sigma({1\over2}^-)$ photoproduction  is
related to the Lagrangian of Eq.(\ref{kns}). Following
Refs.~\cite{hab06,oh08}, we adopt the contact current
\begin{equation}
M_c^\nu=ie g_{KN\Sigma}C^\nu\,
\end{equation}
for $\gamma p\rightarrow K^+\Sigma^0({1\over2}^-)$, where the Lorenz
index $\nu$ couples to that of the photon, and $C^\nu$ is expressed
as Eq.(\ref{cnup}), where $h=1$ is taken. For $\gamma n\rightarrow
K^+\Sigma^-({1\over2}^-)$ process, we adopt the contact current:
\begin{equation}
M_c^\nu=ie\sqrt{2}g_{KN\Sigma}C^\nu\,,
\end{equation}
where $C^\nu$ is expressed as Eq.(\ref{cnun}), where $h=1$ is taken.
With these contact currents, one can check that the total amplitudes
are gauge invariant.

Since previous evidence for the new $\Sigma({1\over2}^-)$ indicates
its mass to be around $\Sigma^*(1385)$~\cite{wu09}, here we assume
its mass be the same as $\Sigma^*(1385)$. The coupling constant
$g_{KN\Sigma}$ and the relevant cut-off parameter $\Lambda_M$ are
unknown parameters, and will be tuned to fit the data.

\section{RESULTS AND DISCUSSIONS}

The effective Lagrangian methods employ hadronic fields as basic
degrees of freedom for the interactions, which introduce many
parameters such as the coupling constants and cutoff parameters.
Many of the parameters are not well constrained, and approximations
such as flavor SU(3) symmetry relation or model predictions are
used. These may bring some uncertainty in the calculation. For
example, in this paper, we consider 3 PDG resonances in s-channel as
in Ref.~\cite{oh08}. Their masses, widths and coupling constants are
not well constrained by experiments. Our choices of the parameters
from the allowed range of PDG~\cite{PDG08} and model
calculations~\cite{capstick92,capstick98} certainly have big
uncertainties, similar as discussed in Ref.~\cite{van09} for the
study of $\gamma n\to K^+\Sigma^-$. However, for the cross sections
and beam asymmetry of the reaction $\gamma n\to
K^+\Sigma^{*-}(1385)\to K^+\Lambda\pi^-$ at $E_\gamma=1.5-2.4$~GeV
reported by the LEPS collaboration ~\cite{leps09}, these 3
resonances mainly contribute to cross sections around
$E_\gamma=1.8$~GeV, and their contributions are small for
$E_\gamma>$2.1~GeV. Even without constraint from other sources,
adjusting these parameters cannot reproduce the measured beam
asymmetry qualitatively. We need to find other ingredients to
describe the cross sections and beam asymmetry data simultaneously.


The measurement by the LEPS Collaboration is limited to forward
angles with $\cos\theta_{\mathrm{c.m.}}\geq 0.6$, where
$\theta_{\mathrm{c.m.}}$ is the forward angle of the produced $K$
meson in the c.m. frame. The theoretical predictions by the original
set of parameters ($h=1$ for the contact term)~\cite{oh08} deviate
from the LEPS results for the linear beam asymmetry~\cite{leps09},
which can be interpreted as
\begin{equation}\label{bma}
A_{beam}={\sigma_\bot -\sigma_{\parallel}\over \sigma_\bot
+\sigma_{\parallel}}\,,
\end{equation}
where $\sigma_\bot$ and $\sigma_{\parallel}$ denote the cross
sections for beam polarization vertical and parallel to the reaction
plane, respectively.

\begin{table*}[tbh]
\begin{tabular}{c|c|c|c|c|c|c|c}
\hline\hline scheme & $h$ & $\Lambda_M$  & $\Gamma_{N^*(2080)}$   &
$\Gamma_{\Delta(1940)}$ &
 $\Gamma_{\Delta(2000)}$ & $g_{KN\Sigma}$~($\Lambda_M$) & $\chi^2$ \\
 \hline

I & 1.0 (fixed) & 0.8~GeV& 0.25~GeV& 0.15~GeV & 0.15~GeV & 1.34~(1.6
GeV)&97
\\
II & 1.11 &  &  &  &  & 0 (fixed) &102
\\\hline
\cite{leps09,oh08} (PDG~\cite{PDG08}) & 1.0 (fixed) & 0.83 & 0.3
($0.12\sim 0.63$)& 0.3 ($0.10\sim 0.78$) & 0.3 ($0.07\sim 0.52$) & 0
(fixed)
&$\sim 180$  \\
\hline\hline
\end{tabular}
\caption{Adjusted parameters for $\gamma n\rightarrow
K^+\Sigma^{*-}$ with two schemes compared with original ones in
Refs.~\cite{leps09,oh08} and PDG range~\cite{PDG08}, and
corresponding $\chi^2$ for 39 data points in Figs.~3 \&~4.}
\label{tab:PAR}
\end{table*}

In this work, we find two possible ingredients to explain the
observed beam asymmetry. One is by including the possible
contribution of $K\Sigma({1\over 2}^-)$ photoproduction, and another
is to assume a different h parameter for the $\gamma n$ reaction
from the $\gamma p$ reaction. To incorporate the new ingredients to
fit both cross sections and beam asymmetry, the minimum set of
parameters to be adjusted from those used in Ref.~\cite{leps09,oh08}
are listed in Table I. For both schemes, the narrower widths for the
three $N^*$ and $\Delta^*$ resonances are used, but still within the
PDG uncertainties~\cite{PDG08}. The reached $\chi^2$ of the two
schemes are also listed. In scheme I, we keep $h=1$ fixed for the
contact term to be identical to the $\gamma p\rightarrow
K^+\Sigma^{*0}$ process, and then include the contribution of
$K\Sigma({1\over 2}^-)$ photoproduction for the process. The
coupling constant of $\Sigma({1\over 2}^-)$ to $KN$ and the
corresponding cut-off parameter are tuned to describe the
experimental data. In scheme II, we do not consider $\Sigma({1\over
2}^-)$ production, and tune parameter $h$ of the contact term to
describe the data.


\begin{figure}\label{fig2}
{\includegraphics[width=0.8\columnwidth]{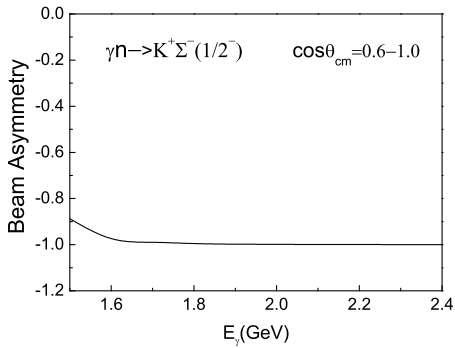}}
 \caption{Linear beam asymmetry for $\gamma +n\rightarrow
K^++\Sigma^-({1\over2}^-)$ with $\cos\theta_{\mathrm{c.m.}}=0.6-1$.}
\end{figure}

In Fig.~2 we show the result of the linear beam asymmetry for
$\gamma n\rightarrow K^+\Sigma^-({1\over2}^-)$ with
$\cos\theta_{\mathrm{c.m.}}\geq 0.6$, {\it i.e.}, the cross sections
in Eq.~(\ref{bma}) are integrated results for
$\cos\theta_{\mathrm{c.m.}}=0.6-1$. The process has a large negative
beam asymmetry that approaches $-1$. This result comes from the fact
that $K^+\Sigma^-({1\over2}^-)$ is produced mainly from the
t-channel kaon exchange process. For the t-channel kaon exchange
process, the photon spin is necessary to be vertical to the reaction
plane due to angular momentum conservation, and such photon
corresponds to the beam (electromagnetic field) with polarization
parallel to the reaction plane, {\it i.e.}, $\sigma_\bot=0$. Thus
the $K^+\Sigma^-({1\over2}^-)$ production contributes negatively to
the beam asymmetry. If a portion of the detected $\Lambda\pi$ stems
from $\Sigma({1\over2}^-)$, the measured beam asymmetry can be
pulled to the negative side, and the observed negative beam
asymmetry may be explained.

\begin{figure}\label{fig3}
{\includegraphics[width=1.08\columnwidth]{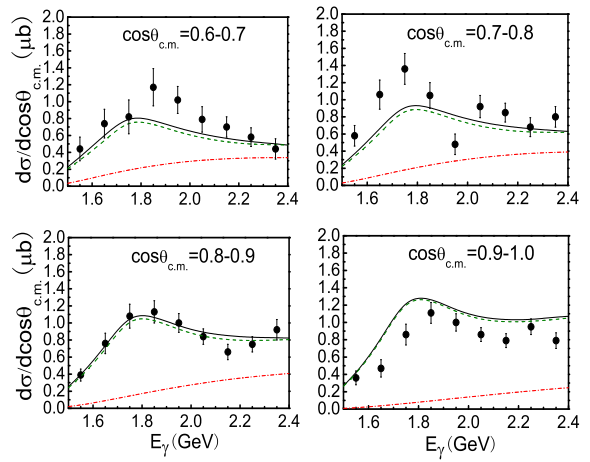}}
 \caption{Differential cross sections for $\gamma +n\rightarrow
K^++\Sigma^{(*)-}$ with four $\cos\theta_{\mathrm{c.m.}}$ intervals.
Theoretical results with scheme I (solid curves) and scheme II
(dashed curves) are compared with the LEPS data~\cite{leps09}. The
dot-dashed curves demonstrate the contributions from $\Sigma({1\over
2}^-)$ production in scheme I.}
\end{figure}

In Fig.~3, the differential cross sections for the process $\gamma
n\to K^+\Sigma^{*-}$ with respect to $\cos\theta_{\mathrm{c.m.}}$
are shown and compared with the LEPS data~\cite{leps09}. The solid
lines are the results of scheme I, and the dashed lines are the
results of scheme II. The contributions from $\Sigma({1\over 2}^-)$
production in scheme I are also shown by the dash-dotted lines.

\begin{figure}\label{fig4}
{\includegraphics[width=0.95\columnwidth]{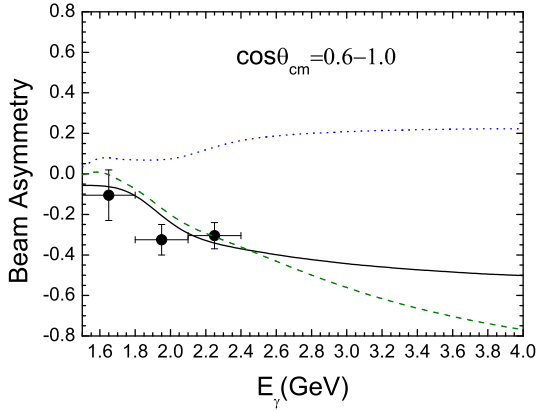}}
 \caption{The linear beam asymmetry for $\gamma +n\rightarrow
K^++\Sigma^{(*)-}$ with $\cos\theta_{\mathrm{c.m.}}$ integrated from
0.6 to 1. Results of scheme I (solid curve), and scheme II (dashed
curve) are compared with the LEPS data~\cite{leps09}. The dotted
curve demonstrate the result with $h=1$ and without $\Sigma({1\over
2}^-)$ contribution.}
\end{figure}

\begin{figure}\label{fig5}
{\includegraphics[width=0.95\columnwidth]{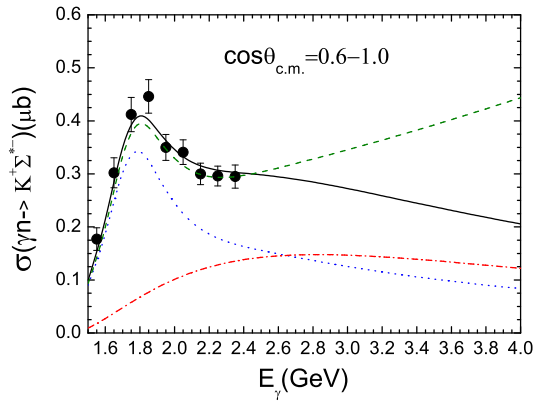}}
 \caption{Integrated cross sections for $\gamma +n\rightarrow
K^++\Sigma^{(*)-}$ with $\cos\theta_{\mathrm{c.m.}}=0.6-1.0$ of
scheme I (solid curve) and scheme II (dashed curve), compared with
the LEPS data~\cite{leps09}. The dotted and dot-dashed curves
demonstrate the contributions from $\Sigma^*({3\over 2}^+)$ and
$\Sigma({1\over 2}^-)$, respectively, in the scheme I. }
\end{figure}

In Fig.~4 and Fig.~5, the results for the linear beam asymmetry and
integrated cross sections of the reaction $\gamma n\rightarrow
K^+\Sigma^{*-}$ are shown and compared with the LEPS
data~\cite{leps09}, respectively. The cross sections are intergrated
results in phase space $\cos\theta_{\mathrm{c.m.}}= 0.6-1.0$. The
solid lines are the results of scheme I, {\it i.e.}, including both
$\Sigma^*({3\over 2}^+)$ production with $h=1$ and the contribution
from $\Sigma({1\over 2}^-)$. The dashed lines are the results of
scheme II, {\it i.e.}, for $\Sigma^*({3\over 2}^+)$ production alone
with $h=1.11$. The dotted lines are the results with the $h=1$ and
without the contribution from $\Sigma({1\over 2}^-)$. The
contribution from $\Sigma({1\over 2}^-)$ production to the cross
section alone is shown by the dash-dotted line in Fig.~5.

From Fig.3-5, one can see that both schemes can well describe the
LEPS data. For scheme I, since the $\Sigma({1\over 2}^-)$ alone
gives a large negative beam asymmetry approaching $-1$ as shown in
Fig.~2, a portion of $\Sigma({1\over 2}^-)$ in this process helps to
explain the observed negative beam asymmetry. At the same time, we
find that tuning the parameter $h$ to a larger value as scheme II
can also have such an effect. The $\chi^2$ for the two schemes are
97 and 102, respectively, compared to the 39 experimental data
points in Figs.~3 \& 4. As a comparison, the corresponding previous
theoretical prediction~\cite{leps09} without including the two new
ingredients give a $\chi^2$ value about 180.

From the comparison of the results with the LEPS data, one can not
tell for sure which one of the two schemes is better. In fact, some
other combinations of $h$ and $g_{KN\Sigma}$, such as $h=1.06$
combined with $g_{KN\Sigma}=1.04$, may also describe the present
data. However, one can see from Fig.~5 that for higher incident
energy, the two schemes give very distinctive predictions. Thus the
schemes and mechanisms can be distinguished when data on reaction
$\gamma n\rightarrow K^+\Sigma^{*-}$ at higher incident energies are
available.

The above results show that while the original theoretical
prediction with $\Sigma^*({3\over 2}^+)$ production alone with $h=1$
fails to reproduce the data for $\gamma n\to
K^+\Sigma^{*-}$~\cite{leps09,oh08}, the data can be well reproduced
either by increasing $h$ to 1.11 or by including an additional
$\Sigma({1\over 2}^-)$ resonance around 1380 MeV. Since the
parameters for the original theoretical prediction come from the
fits to the data on the $\gamma p\to K^+\Sigma^{*0}$ process, we
need to check how the new sets of parameters fit to the $\gamma p\to
K^+\Sigma^{*0}$ data. The results are shown in Fig.~6 for the total
cross sections of the reaction $\gamma p\rightarrow K^+\Sigma^{*0}$
with the above sets of parameters, comparing with the CLAS
data~\cite{clasg06}. Again, the solid line is the result for
$\Sigma^*({3\over 2}^+)$ production with $h=1$ combined with the
contribution from $\Sigma({1\over 2}^-)$ with $g_{KN\Sigma}=1.34$;
the dotted line and the dashed line are the results for
$\Sigma^*({3\over 2}^+)$ production alone with $h=1$ and $h=1.11$,
respectively. The $\Sigma({1\over 2}^-)$ contribution is also shown
by the dash-dotted line with $g_{KN\Sigma}=1.34$. From Fig. 6 one
sees that the combined result of $\Sigma^*({3\over 2}^+)$ production
with $h=1$ and the $\Sigma({1\over 2}^-)$ production (scheme I,
solid line) can be compatible with the CLAS data on the whole, while
the results with $h=1.11$ (dashed line) deviate from the data for
larger incident energy.

\begin{figure}\label{fig6}
{\includegraphics[width=0.95\columnwidth]{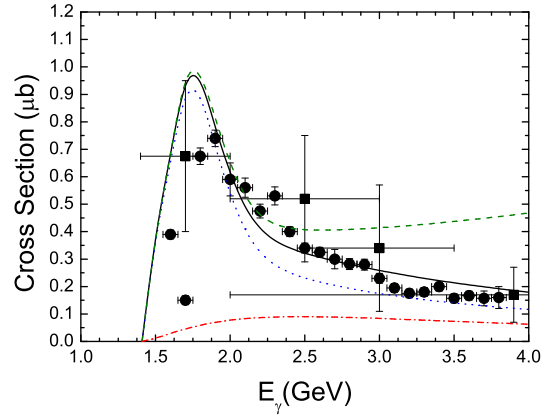}}
 \caption{Total cross sections for $\gamma +p\rightarrow
K^++\Sigma^{(*)0}$ of scheme I (solid curve) and scheme II (dashed
curve), compared with the CLAS preliminary data~\cite{clasg06}. The
dotted and dot-dashed curves demonstrate the contributions from
$\Sigma^*({3\over 2}^+)$ and $\Sigma({1\over 2}^-)$, respectively,
in the scheme I. }
\end{figure}

Our results show that the combined results of $\Sigma^*({3\over
2}^{+})$ and $\Sigma({1\over 2}^-)$ production (solid line) can be
compatible with both the $\gamma n\rightarrow K^+\Sigma^{*-}$
reaction data and the $\gamma p\rightarrow K^+\Sigma^{*0}$ reaction
data with the same set of parameters (scheme I). Without the
$\Sigma({1\over 2}^-)$ contribution, different parameter $h$ need to
be used for $\gamma n\rightarrow K^+\Sigma^{*-}$ and $\gamma
p\rightarrow K^+\Sigma^{*0}$ processes.

The properties and couplings for $\Sigma({1\over 2}^-)$, the contact
term, and the s-channel resonances still have some uncertainties and
need further investigation. More data of broader energy range on
these processes will be helpful to clarify the roles of the contact
term and the $\Sigma({1\over 2}^-)$ resonance.

\section{SUMMARY}

In summery, we study the process of $K\Sigma^*(1385)$
photoproduction from the nucleons in the framework of the effective
Lagrangian method. From recent studies, there is possibly a $\Sigma$
state with $J^p={1\over 2}^-$ near 1380~MeV. We consider the case
that $\Sigma({1\over 2}^-)$ may contribute to the observables of
$K\Sigma^*(1385)$ photoproduction in experiments. Our results show
that the $\Sigma({1\over 2}^-)$ production can give large negative
contribution to beam asymmetry, which helps to explain the large
negative linear beam asymmetry observed by the LEPS experiment. With
a portion of $\Sigma({1\over 2}^-)$, the same set of parameters can
 reproduce both the data of $\gamma n\to K^+\Sigma^{*-}$ from
the LEPS experiment and the data of $\gamma p\to K^+\Sigma^{*0}$
from the CLAS experiment. On the other hand, without including
$\Sigma({1\over 2}^-)$, the present data on $\gamma n\to
K^+\Sigma^{*-}$ can be described by tuning a parameter $h$ in the
contact term, which means different choices of parameter $h$ in the
contact term for $\gamma n$ and $\gamma p$ reactions are needed. To
distinguish the roles of $\Sigma({1\over 2}^-)$ and the contact
term, different predictions of the two schemes are presented for
future experimental study.


\begin{acknowledgments}
We thank Kanzo Nakayama and Yongseok Oh for the helpful discussions.
This project is supported by the National Natural Science Foundation
of China under Grant 10905059, 10875133, 10821063, and by China
Postdoctoral Science Foundation and the Ministry of Science and
Technology of China (2009CB825200).
\end{acknowledgments}

\end{document}